\documentclass[letterpaper,twocolumn,preprintnumbers,amsmath,amssymb,amsfonts,floatfix,nofootinbib,showkeys]{revtex4}
\usepackage{graphicx}
\usepackage{amssymb}
\usepackage{amsmath}
\usepackage{color}
\newif\ifcomment
\newif\ifsoftx
\newif\ifprint
\printtrue
\graphicspath{{./pics/}}
\ifprint
 \usepackage[pdftex,colorlinks,
             linkcolor=red,citecolor=red,anchorcolor=red,
             pagecolor=red,urlcolor=red]{hyperref}
\else
 \usepackage[pdftex,backref,pagebackref,colorlinks,
             linkcolor=blue,citecolor=blue,anchorcolor=blue,
             pagecolor=blue,urlcolor=red]{hyperref}
\fi
\pdfoutput=1        
\pdfcatalog{/PageMode(/UseOutlines)} 
\newcommand {\version}     {\href{http://www.hepforge.org/downloads/tglaubermc}{v2.7}}
\newcommand {\snn}         {\ensuremath{\sqrt{s_{\scriptscriptstyle{{\rm NN}}}}}}
\newcommand {\signn}       {\ensuremath{\sigma_{\scriptscriptstyle{{\rm NN}}}}}
\newcommand {\ep}          {\mbox{$\epsilon_{\rm part}$}}
\newcommand {\erp}         {\mbox{$\epsilon_{\rm RP}$}}
\newcommand {\Ncoll}       {\ensuremath{N_{\rm coll}}}
\newcommand {\Npart}       {\ensuremath{N_{\rm part}}}
\newcommand {\arxiv}[1]    {\href{http://www.arxiv.org/abs/#1}{\mbox{arXiv:#1}}}
\newcommand {\hrefurl}[1]  {\href{#1}{\url{#1}}}
\newcommand {\Ref}[1]      {Ref.~\cite{#1}}

\newcommand {\Fig}[1]      {Fig.~\ref{#1}}
\newcommand {\hide}[1]     {\color{white}#1\color{black}}
\newcommand {\lmyangle}    {\ensuremath{\{}}
\newcommand {\rmyangle}    {\ensuremath{\}}}
\newcommand {\co}[1]       {}
\begin{document}

\title{Improved version of the PHOBOS Glauber Monte Carlo}
\author{C.\ Loizides$^1$, J.\ Nagle$^2$, P.\ Steinberg$^3$}
\affiliation{
$^1$Lawrence Berkeley National Laboratory, Berkeley, California, USA\\
$^2$University of Colorado, Boulder, Colorado, USA\\
$^3$Brookhaven National Laboratory, Upton, NY 11973, USA
}

\begin{abstract}\noindent
``Glauber'' models are used to calculate geometric quantities in the
initial state of heavy ion collisions, such as impact parameter,
number of participating nucleons and initial eccentricity.
Experimental heavy-ion collaborations, in particular at RHIC and LHC, use 
Glauber Model calculations for various geometric observables for
determination of the collision centrality.
In this document, we describe the assumptions inherent to the
approach, and provide an updated implementation (v2) of the Monte Carlo based 
Glauber Model calculation, which originally was used by the PHOBOS collaboration. 
The main improvements w.r.t.\ the earlier version (v1)~\cite{Alver:2008aq} are
the inclusion of Tritium, Helium-3, and Uranium, as well as the treatment
of deformed nuclei and Glauber-Gribov fluctuations of the proton in p+A collisions.
A users' guide (updated to reflect changes in v2) is provided for running 
various calculations.\\

\noindent {\bf Revisions \&\& changes of the arXiv document and code:}\\
 {\bf v1}, 11 Aug 2014: initial document, code v2.0 \\
 {\bf v2}, 20 Dec 2014: updated Table~\ref{tab:awR} with ``p'' and ``PbHN'', code v2.1 (fixes bug for Hulthen) \\
 {\bf v3}, 01 Sep 2016: published version, fixed Eq.~\ref{eq3}, code v2.3 (added $b_{\rm NN}$, and various proton profiles\\\color{white}{\bf v3}, 01 Sep 2016: \color{black}from Pythia, changes in calculation of $\varepsilon_n$) \\
 {\bf v4}, 07 Jun 2017: updated Table~\ref{tab:awR} with ``Xe'' and corrected typo in proton radius, code v2.4\\\color{white}{\bf v4}, 07 Jun 2017: \color{black}(fixed drawing function)\\
 {\bf v5}, 16 Oct 2017: updated Table~\ref{tab:awR} with ``Pb*''\\
 {\bf v6}, 13 Nov 2017: updated Table~\ref{tab:awR} with ``Xes''\\
 {\bf v7}, 01 Mar 2018: split previous Table~1 into two tables: Table~\ref{tab:awR} for spherical and Table~\ref{tab:awR2} for\\\color{white}{\bf v4}, 01 Mar 2018: \color{black} deformed nuclei parameters, added ``Xe2'' and ``Xe2a'' to include deformation of\\\color{white}{\bf v7}, 01 Mar 2018: \color{black} Xenon, code v2.5 (included core/corona calculation, see ``Npart0'')\\
 {\bf v8}, 30 Apr 2018: fixed typo in $Y_{20}$, fixed $\beta_4$ in ``Si2'', added ``Al'', describes new function\\\color{white}{\bf v8}, 25 Apr 2018: \color{black} {\it runAndCalcDens}, code v2.6 (includes fixes and ``runAndCalcDens'')\\
 {\bf v9}, 14 Jan 2019: added Helium-4, Carbon and Oxygen from wavefunctions, unified deuteron\\\color{white}{\bf v9}, 14 Jan 2019: \color{black}treatment, added and new function {\it runAndOutputLemonTree}, updated Table~\ref{tab:awR},\\\color{white}{\bf v9}, 14 Jan 2019: \color{black}latest code \version\ (includes new elements and ``runAndOutputLemonTree'')\\
\end{abstract}
\keywords{Glauber Monte Carlo model, participant, eccentricity, centrality}

\maketitle

\ifsoftx
\else
\section{Introduction}
\fi
\label{sec:intro}
In heavy-ion collisions, initial geometric quantities such as impact
parameter and shape of the collision region cannot be directly
determined experimentally. However, it is possible to relate the
number of observed particles and number of spectator neutrons to the
centrality of the collision. Using the percentile centrality of a
collision, the initial geometric configuration can be estimated with
models assuming the configuration nuclei from nucleons and
the collision process of two nuclei.

These models fall in two main classes. (For a review, see
\Ref{glaubreview}). In the so called ``optical'' Glauber
calculations, a smooth matter density is assumed, typically described
by a Fermi distribution in the radial direction and uniform over solid
angle. In the Monte Carlo~(MC) based models, individual nucleons are
stochastically distributed event-by-event and collision properties are
calculated by averaging over multiple events. As discussed in
\Ref{glaubreview} and \Ref{eccentricity}, the two type of models
lead to mostly similar results for simple quantities such as the number of
participating nucleons~($\Npart$) and impact parameters~($b$), but
give different results in quantities where event-by-event fluctuations
are significant, such as participant frame eccentricity~($\ep$).

In this paper, we discuss several improvements of the Monte Carlo Glauber
calculation originally implemented by the PHOBOS collaboration~\cite{Alver:2008aq}.
The main improvements w.r.t.\ the earlier version (v1) are
the inclusion of Tritium, Helium-3, and Uranium, as well as the treatment
of deformed nuclei and Glauber-Gribov fluctuations of the proton in p+A collisions.
Another widely used implementation of a MC Glauber calculation
can be found in \Ref{Rybczynski:2013yba}.
\ifsoftx
\else
In section~\ref{sec:method}, the MC approach is outlined and the assumptions 
that go into the calculation are introduced. In section~\ref{sec:howto}, 
we discuss the implementation and the tutorial functions provided.
Section~\ref{sec:conclusions} summarizes the paper.
\fi
The latest version~2 code described in this document is \version. 
For further improvements, mainly focusing on nuclear collisions involving Pb at high collision energies, see \Ref{Loizides:2017ack}.

\ifsoftx
\else
\section{The model}\label{sec:method} 
\fi
The MC Glauber model calculation is performed in two
steps. First, the nucleon positions in each nucleus are stochastically
determined. Then, the two nuclei are ``collided'', assuming the
nucleons travel in a straight line along the beam axis (eikonal approximation) 
such that nucleons are tagged as wounded~(participating) or spectator.

\ifsoftx
\else
\subsection{Makeup of nuclei}
\fi
\label{sec:nucleus}
The position of each nucleon in the nucleus is determined according to
a probability density function. 
In a quantum mechanical picture, 
the probability density function can be thought of as the single-particle
probability density and the position as the result of a position
measurement. In the context of the MC Glauber model, one commonly requires
a minimum inter-nucleon separation~($d_{\rm min}$) between the centers of the nucleons
for the determination of the nucleon positions inside a given nucleus.

The probability distribution for spherical nuclei is taken to be uniform in
azimuthal and polar angles. 
The radial probability function is modeled from nuclear charge densities 
extracted in low-energy electron scattering experiments~\cite{atomicdata}. 
The nuclear charge density is usually parameterized by a Fermi
distribution with three parameters
\begin{equation}\label{eq1}
  \rho(r)=\rho_0 \frac{1+w(r/R)^2}{1+\exp(\frac{r-R}{a})}\,,
\end{equation} 
where $\rho_0$ is the nucleon density, $R$ is the nuclear radius, $a$
is the skin depth and $w$ corresponds to deviations from a spherical
shape. The overall normalization ($\rho_0$) is not relevant for this
calculation. Values of the other parameters used for different nuclei
are listed in Table~\ref{tab:awR}.

Exceptions from Eq.~\ref{eq1} are the Deuteron~(${}^{2}$H), Tritium~(H3, ${}^{3}$H), 
Helium-3~(He3, ${}^{3}$He), Helium-4~(He4, ${}^{4}$He), Carbon~(C), Oxygen~(O) and Sulfur (${}^{32}$S) nuclei. 
For Sulfur, a three parameter Gaussian form is used
\begin{equation}\label{eq2}
  \rho(r)=\rho_0 \frac{1+w(r/R)^2}{1+\exp(\frac{r^2-R^2}{a^2})}\,.
\end{equation} 
The values of $R$, $a$ and $w$ for Sulfur are also given in
Table~\ref{tab:awR}. For Deuteron, three options are supported
with the $3^{\text{rd}}$ option being the one used in PHOBOS analyses:
\begin{enumerate}
\item The three parameter Fermi form with the values given in Table~\ref{tab:awR}.
\item The Hulth\'en form
\begin{equation}\label{eq3}
  \rho(r')=\rho_0 \left(\frac{e^{-ar'}-e^{-br'}}{r'}\right)^2,
\end{equation} 
where $a=0.228$~fm$^{-1}$ and $b=1.177$~fm$^{-1}$, and $r'$ denotes the distance between the proton and neutron, i.e.\ $r=r'/2$ \cite{Hulthen, deuteronpars, Adler:2006xd}.
\item The proton is taken from the Hulth\'en form given
above with the neutron placed opposite to it. 
\end{enumerate}
For ${}^{3}$H and ${}^{3}$He, the configurations 
were computed (and stored in a database) from Green's function MC calculations 
using the AV18 + UIX model interactions, which correctly sample the
position of the three nucleons, including correlations, as in \Ref{Nagle:2013lja}.
Similarly, results of wavefunction-based calculations described in \Ref{Lim:2018huo} were made available for He-4, Carbon and Oxygen.
Finally, note that we also provide the rescaled values for the radius and skin depth 
of  Cu~(``CuHN''), Au~(``AuHN'') and Pb~(``PbHN'') to take into account the finite nucleon profile 
as derived in \cite{Hirano:2009ah} and \cite{Shou:2014eya}.
For deformed nuclei we use
\begin{equation}\label{eq:deformed}
  \rho(x,y,z)=\rho_0 \frac{1}{1+\exp\frac{\left(r-R(1+\beta_2 Y_{20} +\beta_4 Y_{40})\right)}{a}}\,,
\end{equation} 
where $Y_{20}=\sqrt{\frac{5}{16\pi}}(3{\rm cos}^2(\theta)-1)$, 
$Y_{40}=\frac{3}{16\sqrt{\pi}}(35{\rm cos}^4(\theta)-30 {\rm cos}^2(\theta)+3)$ with the
deformation parameters $\beta_2$ and $\beta_4$ taken from Ref.~\cite{atomicdata2}. 
The values used for different nuclei~(Si, Cu, Au and U) are listed in Table~\ref{tab:awR2}.

\begin{table}[t!]
\begin{center}
  \begin{tabular}{l|c|c|c}
    Name           & R~[fm]        &  a~[fm]       & w \\
    \hline
    p\footnote{An exponential $\exp{-r/R}$ is used with $R=0.234$~fm based on the rms charge radius of the proton~\cite{hofstadter}. Also single (``pg'') and double Gaussian (``pdg'') profiles are implemented~\cite{Corke:2011yy}. See code.}    
                   & 0.234         &               & \\
    d\footnote{The Hulthen form should be used, with the proton and neutron constrained to be opposite of each other. The parameters are given in the text. For the other options, read the code.}
                   & 0.01\hide{0}  & 0.5882        & \hide{-}0\hide{.0000}       \\
    H3\footnote{The configurations were obtained using wave function calculations as in \Ref{Nagle:2013lja} and \Ref{Lim:2018huo}.} &&&\\
    He3$^{c}$&&&\\
    He4$^{c}$&&&\\
    C$^{c}$&&&\\
    ${}^{16}$O$^{c}$&&&\\
    ${}^{16}$Opar  & 2.608         & 0.513\hide{0} & -0.51\hide{00}              \\
    ${}^{16}$Oho   & 1.833         & 1.544\hide{0} & \hide{-}0\hide{.0000}       \\
    ${}^{28}$Si    & 3.34\hide{0}  & 0.580\hide{0} & -0.233\hide{0}              \\
    ${}^{32}$S     & 2.54\hide{0}  & 2.191\hide{0} & \hide{-}0.16\hide{00}       \\
    ${}^{40}$Ar    & 3.53\hide{0}  & 0.542\hide{0} & \hide{-}0\hide{.0000}       \\
    ${}^{40}$Ca    & 3.766         & 0.586\hide{0} & -0.161\hide{0}              \\
    ${}^{58}$Ni    & 4.309         & 0.517\hide{0} & -0.1308                     \\
    ${}^{63}$Cu    & 4.2\hide{00}  & 0.596\hide{0} & \hide{-}0\hide{.0000}       \\
    ${}^{63}$CuHN  & 4.28\hide{0}  & 0.5\hide{000} & \hide{-}0\hide{.0000}       \\
    ${}^{129}$Xe\footnote{Parameters for $^{132}$Xe $R=5.4\pm0.1$ and $a=0.61^{+0.07}_{-0.09}$ fm from \Ref{Tsukada:2017llu} were used. 
                         The radius is scaled down by $0.99 = (129/132)^{1/3}$ and $a$ was reduced by $0.02$ fm 
                         to symmetrize the uncertainty and to approximate the smaller neutron skin of $^{129}$Xe.}
                  & 5.36\hide{0}  & 0.59\hide{00} & \hide{-}0\hide{.0000}       \\
    ${}^{129}$Xes\footnote{Parameters for Sb~(Antimony) $R=5.32$ and $a=0.57$ fm from \Ref{atomicdata} were used. 
                          Natural Sb has two isotopes: $A=121$ ($57.2$\%) and $A=123$ ($42.8$\%). 
                          The radius of Sb is scaled up by $1.019 = (129/122)^{1/3}$ using the average of $A\approx122$.
                          The resulting parameters are consistent with those obtained from $^{132}$Xe.}
                  & 5.42\hide{0}  & 0.57\hide{00} & \hide{-}0\hide{.0000}       \\
    ${}^{186}$W    & 6.58\hide{0}  & 0.480\hide{0} & \hide{-}0\hide{.0000}       \\
    ${}^{197}$Au   & 6.38\hide{0}  & 0.535\hide{0} & \hide{-}0\hide{.0000}       \\
    ${}^{197}$AuHN & 6.42\hide{0}  & 0.44\hide{00} & \hide{-}0\hide{.0000}       \\
    ${}^{207}$Pb\footnote{These values are usually also used for ${}^{208}$Pb, since the Bessel-Fourier coefficients 
                         for the two nuclei are similar~\cite{atomicdata}. Indeed, an earlier publication~\cite{DeJager:1974liz}
                         gives $R=6.624$ and $a=0.549$~fm for ${}^{208}$Pb~(``Pb*''), consistent within uncertainties with the values for ${}^{207}$Pb.}
                  & 6.62\hide{0}  & 0.546\hide{0} & \hide{-}0\hide{.0000}       \\
    ${}^{207}$PbHN & 6.65\hide{0}  & 0.46\hide{00} & \hide{-}0\hide{.0000}       \\
    ${}^{208}$Pb*  & 6.624         & 0.549\hide{0} & \hide{-}0\hide{.0000}       \\
  \end{tabular}
  \caption{\label{tab:awR}\protect Nuclear charge density parameters for different nuclei taken from \Ref{atomicdata} for equations~\ref{eq1}, \ref{eq2} and \ref{eq3}.
            All implemented configurations can be found in the code in the {\tt TGlauNucleus::Lookup} function.}
\end{center}  
\end{table}

\begin{table}[t!]
\begin{center}
  \begin{tabular}{l|c|c|c|c|c}
    Name            & R~[fm]        &  a~[fm]      & w & $\beta_2$ & $\beta_4$    \\
    \hline
    ${}^{27}$Al     & 3.34\hide{0}  & 0.580\hide{0} & 0 & -0.448        & \hide{-}0.239 \\
    ${}^{28}$Si2    & 3.34\hide{0}  & 0.580\hide{0} & -0.233 & -0.478        & \hide{-}0.250 \\
    ${}^{63}$Cu2    & 4.2\hide{00}  & 0.596\hide{0} & 0 & \hide{-}0.162 & -0.006        \\
    ${}^{129}$Xe2\footnote{Same parameters as for Xe with $\beta_2$ and $\beta_4$ from \Ref{xedef}.}
                  & 5.36\hide{0}  & 0.59\hide{00}  & 0 & 0.161         & -0.003        \\
    ${}^{129}$Xe2a\footnote{Same parameters as for Xe with $\beta_2=0.18\pm0.02$ from interpolation 
                           between measured deformation parameters for the even-A Xe isotopes~\cite{xedef2}, and $\beta_4$ set to zero.}
                  & 5.36\hide{0}  & 0.59\hide{00}  & 0 & 0.18          &  \hide{-}0    \\
    ${}^{197}$Au2   & 6.38\hide{0}  & 0.535\hide{0} & 0 & -0.131        & -0.031        \\
    ${}^{238}$U2    & 6.67\hide{0}  & 0.44\hide{00} & 0 & \hide{-}0.280 & \hide{-}0.093 \\
    ${}^{238}$U\footnote{Implementation as done in \Ref{Heinz:2004ir}.} & 6.67\hide{0}  & 0.44\hide{00} & 0 & \hide{-}0.280 & \hide{-}0.093 \\
  \end{tabular}
  \caption{\label{tab:awR2}\protect Nuclear charge density parameters for deformed nuclei taken from \Ref{atomicdata2}.
            The ``2'' for Si2, Cu2, Au2 and U2 refers to the use of Eq.~\ref{eq:deformed}.
            All implemented configurations can be found in the code in the {\tt TGlauNucleus::Lookup} function.}
\end{center}  
\end{table}

\begin{figure}[t]
\begin{center}
   \includegraphics[width=80mm]{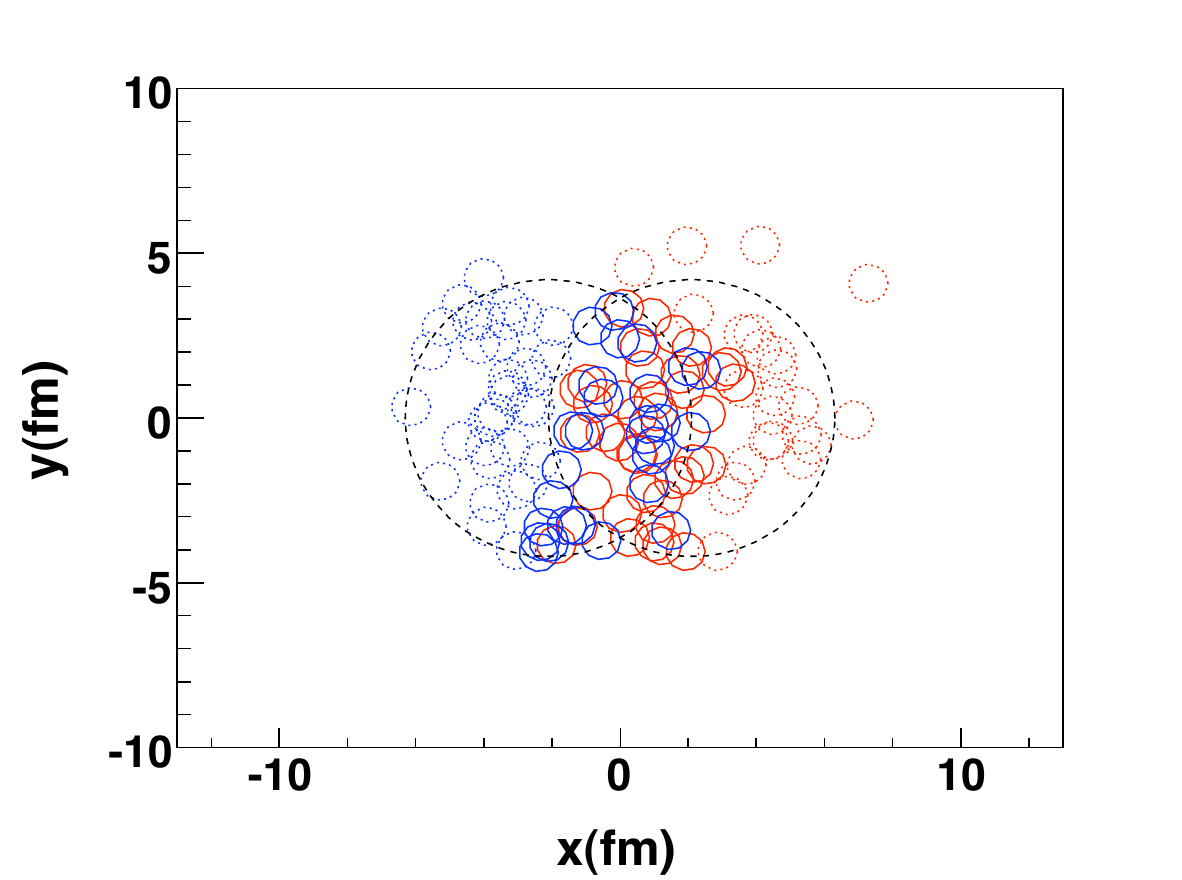}
   \includegraphics[width=80mm]{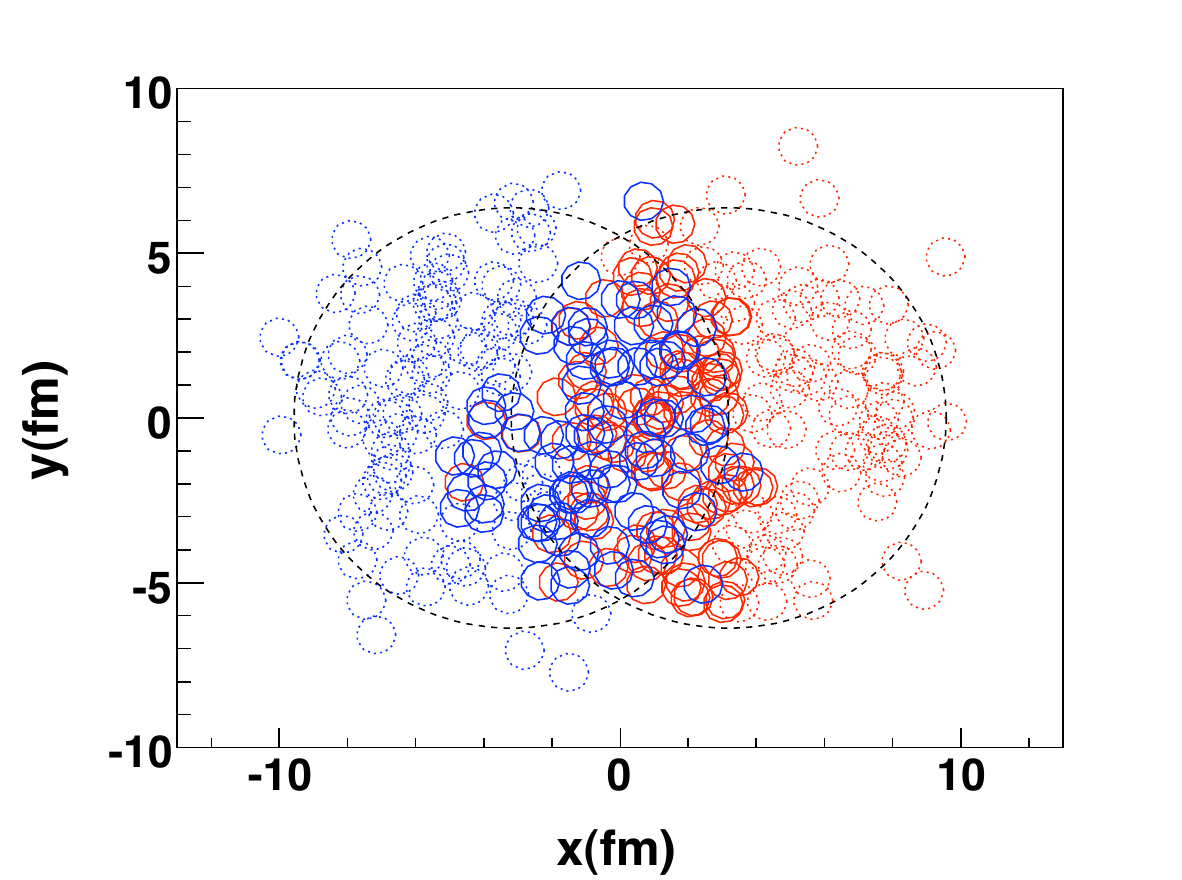}
   \includegraphics[width=80mm]{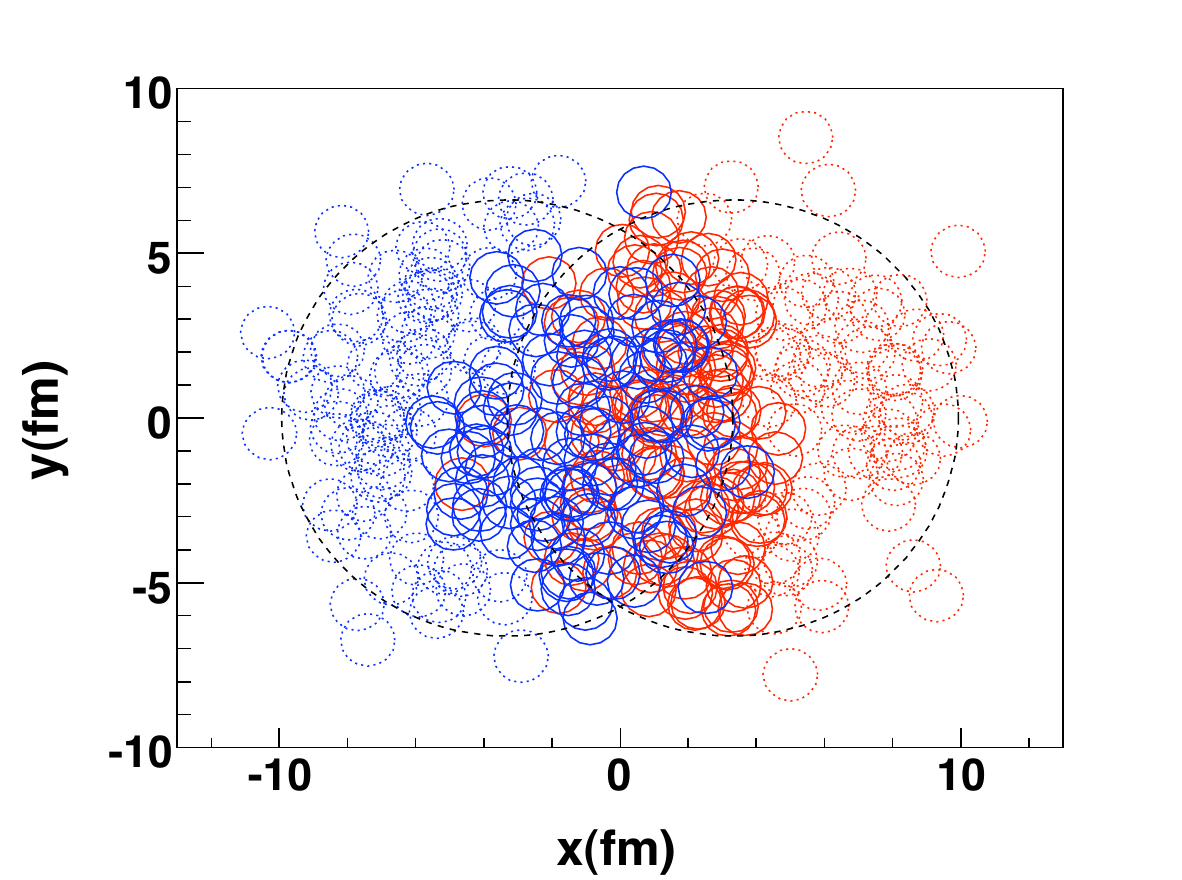}
   \caption{\label{fig:testplots}Typical events for Cu+Cu~(top panel), Au+Au~(middle panel), 
            and Pb+Pb~(lower panel) collisions, the first two performed at RHIC energies 
            and the latter at the LHC. Wounded nucleons~(participants) are indicated as 
            solid circles, while spectators are dotted circles.}
\end{center}
\end{figure}

\ifsoftx
\else
\subsection{Collision Process}
\fi
\label{sec:coll}
The impact parameter of the collision is chosen randomly from a
distribution ${\rm d}N/{\rm d}b \propto b$ up to some large maximum
$b_{\rm max}$ with \mbox{$b_{\rm max}\simeq20\,$fm$>2R_{A}$}. 
The centers of the nuclei are calculated and shifted to $(-b/2,0,0)$ 
and $(b/2,0,0)$. The reaction plane, defined by the impact parameter 
and the beam direction, is given by the $x$- and $z$-axes, while 
the transverse plane is given by the $x$- and $y$-axes.
It is assumed that the nucleons move
along a straight trajectory along the beam axis.
The longitudinal coordinate does not play a role in the calculation.

The inelastic nucleon-nucleon cross section ($\signn$), which is
only a function of the collision energy, is extracted from p+p
collisions. A value of $\signn=42$~mb is used at top RHIC energy of 
$\snn=200$~GeV, while at the LHC $\signn=64$~mb for $\snn=2.76$~TeV
and $\signn=70$~mb for $\snn=5.02$~TeV are used~\cite{glauberlhc}. 
When estimating the systematic  uncertainties on calculated quantities 
like $\Npart$ one varies the $\signn$~(along with the other parameters) 
by typically $\pm$3~mb and $\pm$5~mb at RHIC and LHC, respectively.

The ``ball diameter'' defined as
\begin{equation}
  D = \sqrt{\signn/\pi}
\end{equation}
parameterizes the interaction strength of two nucleons:
Two nucleons from different nuclei are assumed to collide if their
relative transverse distance is less than the ball diameter. 
If no such nucleon--nucleon collision is registered for any pair of nucleons,
then no nucleus--nucleus collision occurred. Counters for determination of the 
total~(geometric) cross section are updated accordingly.

\ifsoftx
\else
\section{Users' guide}
\fi
\label{sec:howto}
The MC Glauber code works within the ROOT framework (ROOT 4.00/08 or
higher~\cite{root}). The code is contained in the macro 
{\tt runglauber\_vX.Y.C}~\cite{glaucode}. The version described here is \version.
Three classes, {\tt TGlauNucleon}, {\tt TGlauNucleus} and {\tt TGlauberMC} 
and three functions~(macros) {\tt runAndSaveNtuple()}, {\tt runAndSaveNucleons()},
and {\tt runAndSmearNtuple()} are defined in the provided macro. 
Note that for generating events with ${}^{3}$H or ${}^{3}$He you will need the text
files called ``h3\_plaintext.dat'' or ``he3\_plaintext.dat'' in the current path.
While the functionality is essentially complete 
for known applications of the Glauber approach, users are encouraged to write their
own functions to access results of the Glauber simulation or to modify the code.

The main classes are the following:
\begin{itemize}
\item{\tt TGlauNucleon} is used to store information about a single nucleon. 
The stored quantities are the position of the nucleon, the number of binary collisions 
that the nucleon has had and which nucleus the nucleon is in, ``A'' or ``B''. 
For every simulated event, the user can obtain an array containing all nucleons~(via {\tt 
TGlauberMC::GetNucleons()}).
\item{\tt TGlauNucleus} is used to generate and store information about a 
single nucleus. The user is not expected to interact with this class.
\item{\tt TGlauberMC} is the main steering class used to generate events and 
calculate event-by-event quantities such as the number of participating nucleons. 
\end{itemize}

\begin{figure}[t]
\begin{center}
   \includegraphics[width=80mm]{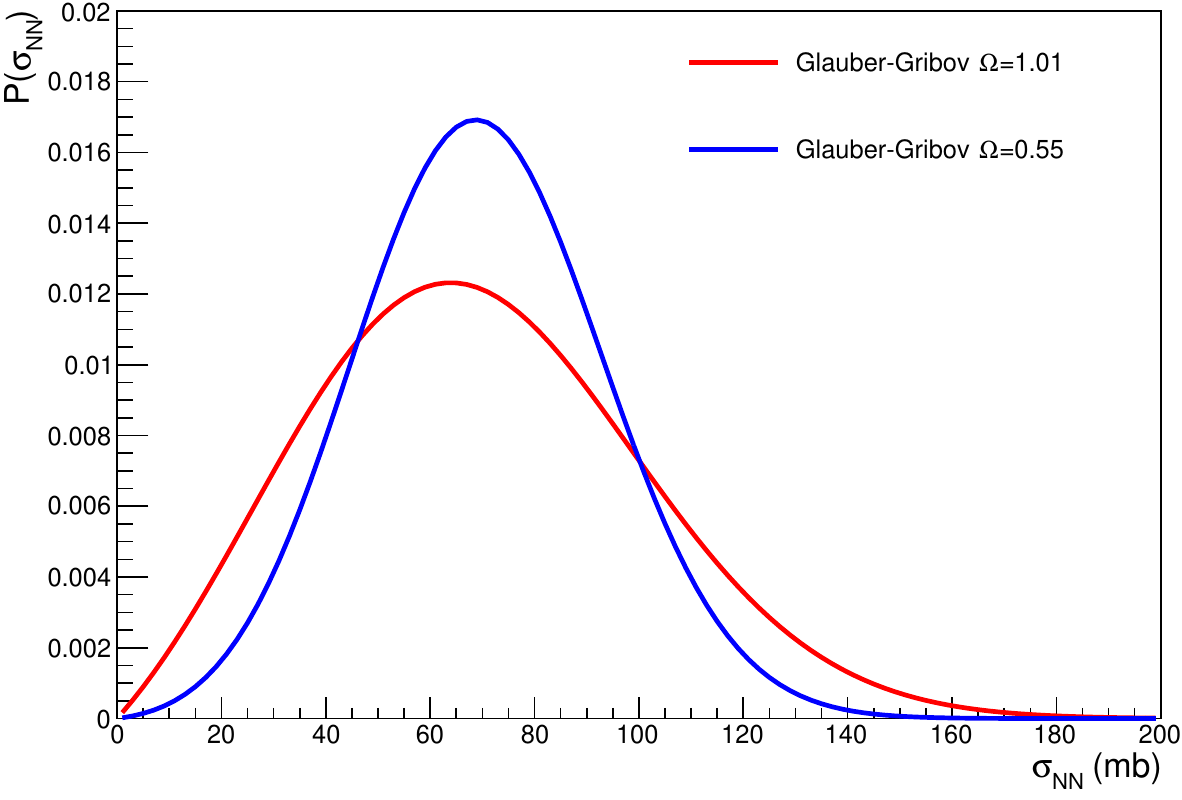}
   \includegraphics[width=80mm]{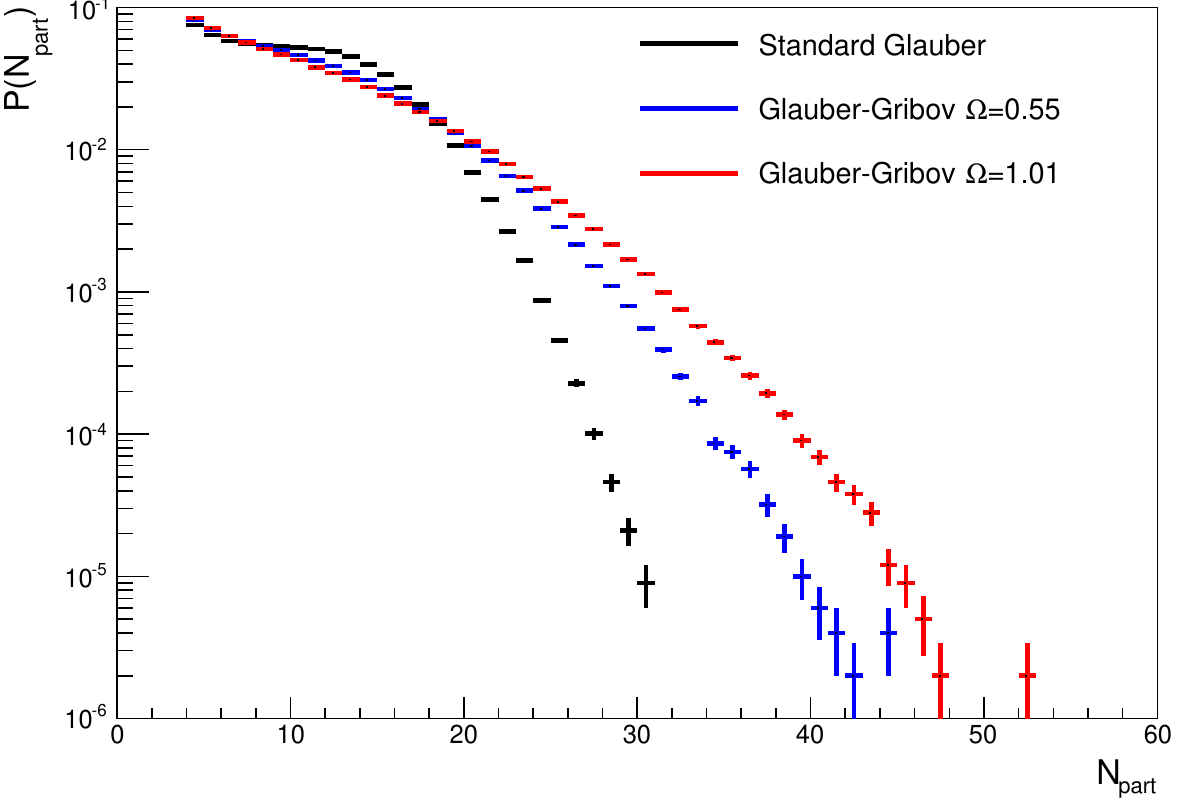}
   \caption{\label{fig:gg} Distribution of $\signn$ for mean $70$~mb with $\Omega=0.55$ and $1.01$ (top panel)
            and resulting $\Npart$ distributions calculated with the Glauber-Gribov extension of the MC Glauber model 
            compared to the standard MC Glauber for p+Pb collisions at the LHC (bottom panel).}
\end{center}
\end{figure}

The steering class {\tt TGlauberMC} has one constructor
\begin{verbatim}
TGlauberMC::TGlauberMC(Text_t* NA, 
                       Text_t* NB, 
                       Double_t xsect,
                       Double_t xsectsigma)
\end{verbatim}
where {\tt NA} and {\tt NB} are the names of the colliding nuclei
and {\tt xsect} and {\tt xsectsigma} are the nucleon-nucleon cross section mean and width 
given in mb. The defined nuclei names are given in Tables~\ref{tab:awR} and \ref{tab:awR2}. 
For Deuteron, the names ``d'', ``dhh'' and ``dh'' correspond to the three options described 
in section~\ref{sec:nucleus} respectively. Units are generally given in fm for distances, while in mb 
for cross sections.
If a finite value ($>0$) for the width of the cross section is given, the calculation will use
the Glauber-Gribov model for fluctuations of the proton. 
This option was originally only implemented for p+A collisions, but can also be enabled for A+A
collisions using the member function ``SetCalcAAGG''.
In this case, following~\cite{Alvioli:2013vk} as described in~\cite{Aad:2015zza} 
$\signn$ will be distributed according to $1/\lambda \, P(\sigma/\lambda)$ with
\begin{equation}
P = \rho \frac{\sigma}{\sigma+\sigma_0} \exp^{-\left(\frac{\sigma-\sigma_0}{\sigma_0\,\Omega}\right)^2}\,
\end{equation}
where $\sigma_0$ denotes the mean $\signn$ and $\Omega$ the width. The normalization $\rho$ and
the rescaling parameter $\lambda=\sigma_0/\left<\sigma\right>$ are computed from the provided input~(mean $\signn$ and $\Omega$)
making use of $\int \sigma P{\rm d}\sigma / \int P{\rm d}\sigma=\sigma_0$ with $P=\rho\sigma_0 \frac{x}{x+1} \exp^{-\left(\frac{x-1}{\Omega}\right)^2}$ for $x=\sigma/\sigma_0$.
Example distributions for p+Pb collisions at the LHC are given in \Fig{fig:gg}.

\ifsoftx
\else
\subsection{Running the Code}
\fi
To generate Au+Au collisions at \mbox{$\snn=200$} GeV~($\signn=42$~mb) one has to
construct a {\tt TGlauberMC} object by issuing at the ``root'' prompt the commands:
\begin{verbatim}
root [0] gSystem->Load("libMathMore")
root [1] .L runglauber_X.Y.C+
root [2] TGlauberMC glauber("Au","Au",42);
\end{verbatim}
where the first ROOT command loads the needed library {\tt libMathMore}, the second
compiles, links and loads the compiled macro~(where you have to replace X.Y with the 
current version number of the code, for example~\version)
including the Glauber code as explained in chapter~2 of the ROOT users' guide, 
while the third commands sets up the Glauber simulation for Au+Au collisions
at $\snn=200$~GeV.

Events can be generated interactively using the two functions
\begin{itemize} 
\item{\tt TGlauberMC::NextEvent(Double\_t bgen)}, which is used to run an event 
at a specified impact parameter, or over a range of impact parameters~(if {\tt bgen=-1}, 
the default value) as described in section~\ref{sec:coll}).
\item{\tt TGlauberMC::Run(Int\_t nevents)} which is used to run a large event sample 
by invoking {\tt NextEvent} many times.
\end{itemize}

Other important public member functions are:
\begin{itemize}
\item {\tt TGlauberMC::SetMinDistance(Double\_t d)}, which is used to set
minimum nucleon separation within a nucleus, $d_{\rm min}$ (default is $0.4$~fm)
\item {\tt TGlauberMC::SetBmin(Double\_t bmin)} and {\tt TGlauberMC::SetBmax(Double\_t bmax)}, 
which can be used to set the range of impact parameter values generated in {\tt Run()}.
\item {\tt TGlauberMC::GetTotXSect()}
which returns the total nucleus-nucleus cross section, calculated when
the function {\tt Run()} is called.
\item {\tt TGlauberMC::Draw()} which draws the current event in the current pad.
\end{itemize}

\subsection{Provided functions}\label{sec:provfunc}
Four functions~(macros) are provided to demonstrate how to run the model:

\subsubsection{runAndSaveNtuple}
The macro {\tt runAndSaveNtuple()} generates a number of Monte Carlo events and
saves some event-by-event quantities. It takes as parameters, the
number of events to be generated, the collision system, the
nucleon-nucleon cross section, its width to simulate Glauber-Gribov
fluctuations (for p+A), the minimum separation distance 
and the output file name. It creates and stores an ntuple in the 
output file with the following event-by-event quantities:
\begin{itemize}
\item{\tt Npart}:       Number of participating nucleons.
\item{\tt Npart0}:      Number of singly-wounded participating nucleons.
\item{\tt Ncoll}:       Number of binary collisions.
\item{\tt B}:           Generated impact parameter.
\item{\tt BNN}:         Mean nucleon--nucleon impact parameter
\item{\tt MeanX}:       Mean of $x$ for wounded nucleons, $\lmyangle x \rmyangle$.
\item{\tt MeanY}:       Mean of $y$ for wounded nucleons, $\lmyangle y \rmyangle$.
\item{\tt MeanX2}:      Mean of $x^2$ for wounded nucleons, $\lmyangle x^2 \rmyangle$.
\item{\tt MeanY2}:      Mean of $y^2$ for wounded nucleons, $\lmyangle y^2 \rmyangle$.
\item{\tt MeanXY}:      Mean of $xy$ for wounded nucleons, $\lmyangle xy \rmyangle$.
\item{\tt VarX}:        Variance of $x$ for wounded nucleons, $\sigma_x^2$.
\item{\tt VarY}:        Variance of $y$ for wounded nucleons, $\sigma_y^2$.
\item{\tt VarXY}:       Covariance of $x$ and $y$ for wounded nucleons, 
                        $\sigma_{xy} \equiv \lmyangle xy \rmyangle - \lmyangle x \rmyangle \lmyangle y \rmyangle$.
\item{\tt MeanXSystem}: Mean of $x$ for all nucleons.
\item{\tt MeanYSystem}: Mean of $y$ for all nucleons.
\item{\tt MeanXA}:      Mean of $x$ for nucleons in nucleus A.
\item{\tt MeanYA}:      Mean of $y$ for nucleons in nucleus A.
\item{\tt MeanXB}:      Mean of $x$ for nucleons in nucleus B.
\item{\tt MeanYB}:      Mean of $y$ for nucleons in nucleus B.
\item{\tt NpartA}:      $\Npart$ from nucleus A.
\item{\tt NpartB}:      $\Npart$ from nucleus B.
\item{\tt PhiA}:        Azimuthal angle for nucleus A.
\item{\tt ThetaA}:      Polar angle for nucleus A.
\item{\tt PhiB}:        Azimuthal angle for nucleus B.
\item{\tt ThetaB}:      Polar angle for nucleus B.
\item{\tt PsiN}:        Event plane angle of Nth harmonic. 
\item{\tt EccN}:        Participant eccentricity for Nth harmonic. 
\end{itemize}

It is important to note that for each of these event-by-event quantities a ``getter'' function
is implemented providing the users the option to write their own event loop~(using {\tt 
TGlauberMC::NextEvent()}).

\subsubsection{runAndSaveNucleons}
The macro {\tt runAndSaveNucleons()} generates a number of Monte Carlo
events and saves an array of {\tt TGlauNucleon} objects for each event.  
It is also possible to use this function to print out the values stored
in the nucleons by setting the verbosity parameter.
The function takes as parameters the number of events to be generated, 
the collision system, the nucleon-nucleon cross section, the minimum separation 
distance, the verbosity flag, minimum and maximum impact parameter and the output file name.

\subsubsection{runAndSmearNtuple}
The macro {\tt runAndSmearNtuple()} generates a number of Monte Carlo events and
saves a reduced set of event-by-event quantities, where the position of
the nucleon is smeared similar to what is done in Ref.~\cite{Nagle:2013lja}.
It takes the same arguments as the {\tt runAndSaveNtuple()} macro.  

\subsubsection{runAndCalcDens}
The macro {\tt runAndCalcDens()} generates a number of Monte Carlo events and
saves for every event a normalized density profile in $x$ and $y$ calculated
from smeared participant and binary collision positions with a relative weight.
As suitable input for hydrodynamical calculations, a scale
factor needs to be applied to adjust to the multiplicity as desribed in Ref.~\cite{Nagle:2013lja}.
The function takes as parameters the number of events to be generated, the
weight, the collision system, the nucleon-nucleon cross section, the minimum separation 
distance, and the output file name.

\subsubsection{runAndOutputLemonTree}
The macro {\tt runAndOutputLemonTree()} generates a number of Monte Carlo events and
saves the output in the ``Lemon TTree'' format needed for the IP-Jazma~\cite{Nagle:2018ybc}.
It takes the same arguments as the {\tt runAndSaveNtuple()}, and optionally can produce a grid with
the energy density distribution.

\begin{figure}[t]
\begin{center}
  \includegraphics[width=80mm]{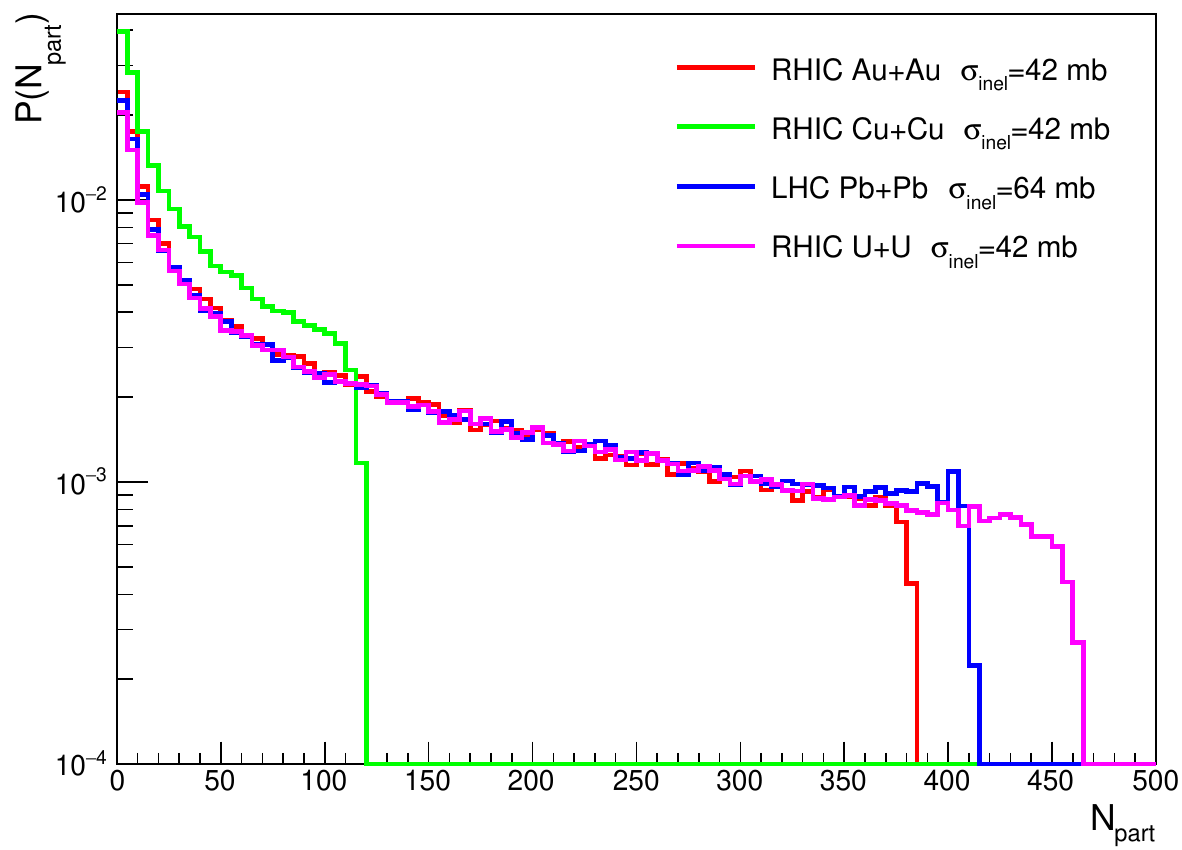}
  \includegraphics[width=80mm]{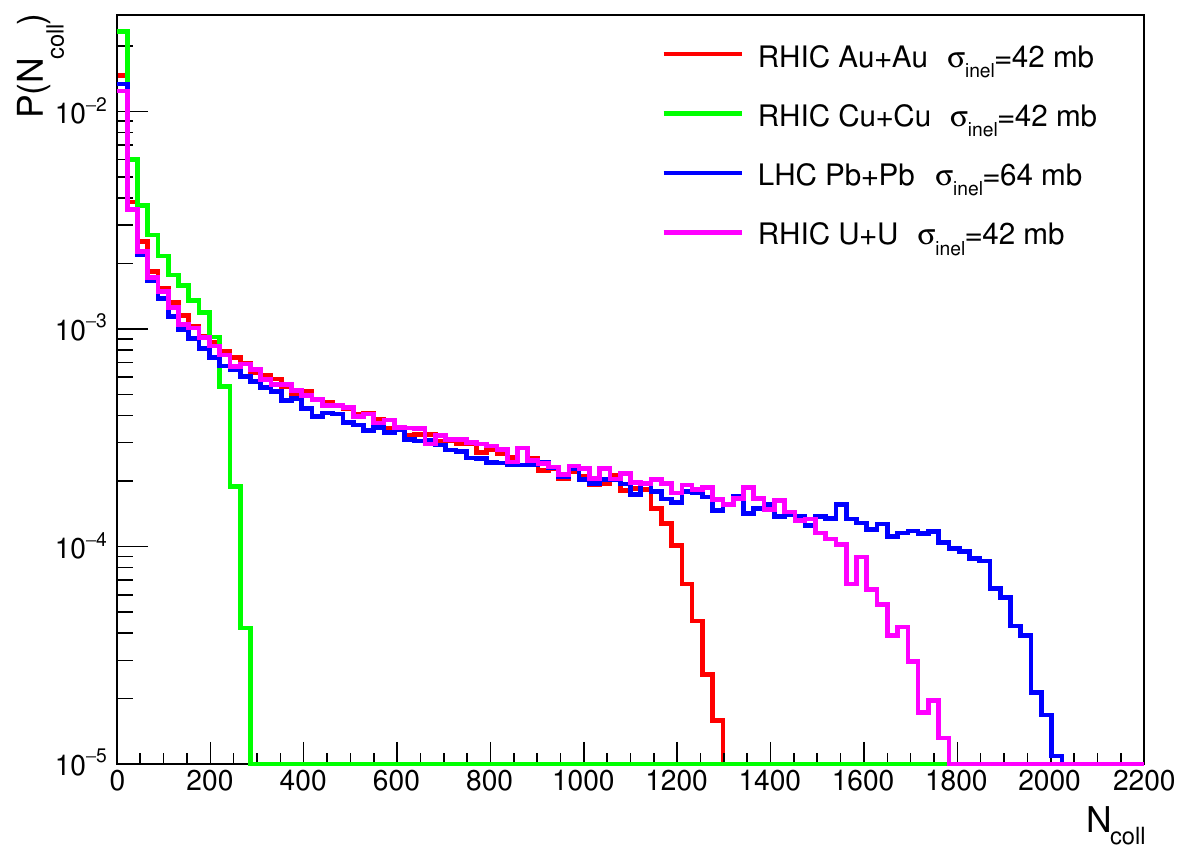}
  \caption{\label{fig:npcplots}Distributions of $\Npart$ (top panel) and $\Ncoll$ (bottom panel) for 
           Cu+Cu, Au+Au and U+U collisions at RHIC, and Pb+Pb collisions at the LHC.}
\end{center}
\end{figure}
\begin{figure}[t]
\begin{center}
  \includegraphics[width=90mm]{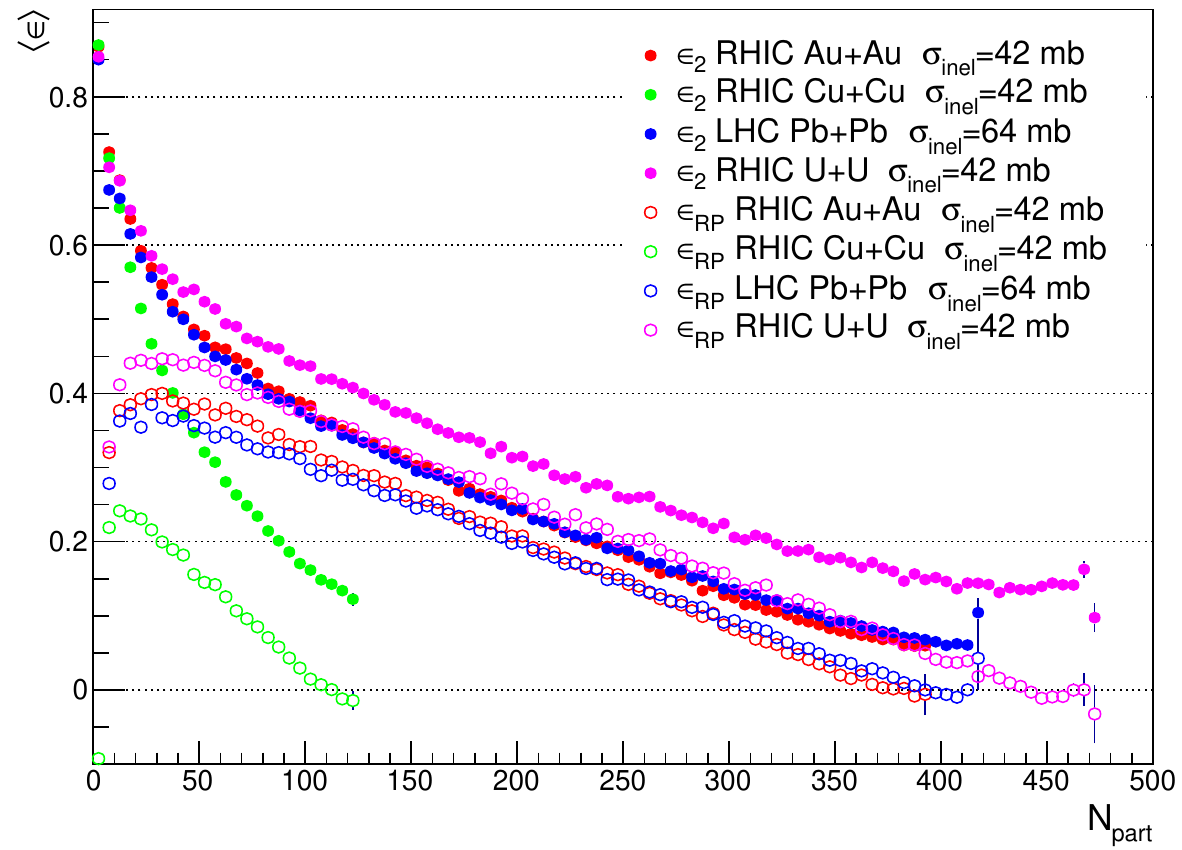}
  \caption{\label{fig:eccplot}Reaction-plane eccentricity $\erp$~(open symbols) and participant-plane eccentricity
           $\ep$~(closed symbols) as a function of $\Npart$ for Cu+Cu, Au+Au and U+U collisions at RHIC and 
           Pb+Pb collisions at the LHC.}
\end{center}
\end{figure}

\ifsoftx
\else
\subsection{Sample Results}
\fi
To provide example applications of this code, 100k events were generated for Cu+Cu, Au+Au and U+U 
at RHIC energies~($\snn=42$~mb), and Pb+Pb at LHC beam energy~($\snn=64$~mb) using the 
{\tt runAndSaveNtuple()} function. 
The resulting ntuples were used to plot the distributions 
of $\Npart$ and $\Ncoll$, shown in \Fig{fig:npcplots}. 

Using the event-by-event
quantities, one can construct combinations of moments like~\cite{eccentricity}:
\begin{itemize}
\item{Reaction-plane eccentricity $\erp$}
\begin{equation}
   \erp = \frac{\text{VarY}-\text{VarX}}{\text{VarY}+\text{VarX}}
\end{equation}
\item{Participant-plane eccentricity $\ep$}
\begin{equation}
   \ep = \frac{\sqrt{(\text{VarY}-\text{VarX})^2+4\text{VarXY}^2}}{\text{VarY}+\text{VarX}}
\end{equation}
\end{itemize}
which are shown in \Fig{fig:eccplot} for different systems.


\ifsoftx
\else
\section{Summary}
\fi
\label{sec:conclusions}
Glauber Model calculations are commonly used by heavy-ion physics 
experiments to study the initial state configurations of nuclear matter.
This document describes the updated implementation (v2) of the Monte Carlo 
based Glauber Model calculation, which originally was used by the 
PHOBOS collaboration. 
The main improvement w.r.t.\ the earlier version~\cite{Alver:2008aq} are
the inclusion of Tritium, Helium-3, and Uranium, as well as the treatment
of deformed nuclei and Glauber-Gribov fluctuations of the proton in p+A collisions.
The code, accessible online, can be used within user code or in a standalone 
mode allowing analysis of various distributions. 
The authors welcome comments on the code and suggestions on how to make
it more useful to both experimentalists and theorists.

 
\end{document}